# Caractérisation des mécanismes d'endommagement des tubes CVI-SiC/SiC par tomographie X

*Characterization of damage mechanisms in CVI-SiC/SiC composite tubes by X-ray tomography*

Yang Chen[1,2], Lionel Gélébart[1], Camille Chateau[2], Michel Bornert[2], Andrew King[3], Patrick Aimediue[2] et Cédric Sauder[1]

1 : DEN-Service de Recherches Métallurgiques Appliquées
CEA, Université Paris-Saclay
CEA Saclay, F- 91191 Gif-sur-Yvette
e-mail : yang.chen@cea.fr; lionel.gelebart@cea.fr; cedric.sauder@cea.fr

2 : Laboratoire Navier (UMR 8205)
CNRS, ENPC, IFSTTAR, Université Paris-Est
F- 77455 Marne-la-Vallée
e-mail : camille.chateau@enpc.fr; michel.bornert@enpc.fr; patrick.aimedieu@enpc.fr

3 : Synchrotron SOLEIL
F- 91192, St-Aubin
e-mail : king@synchrotron-soleil.fr

## Résumé

Les tubes composites SiC/SiC sont étudiés comme des matériaux pour le gainage du combustible nucléaire. La compréhension exhaustive des mécanismes d'endommagement de ce matériau nécessite une étude volumique à la fois expérimentale et numérique. Des essais de traction in situ ont été effectués sous la tomographie de rayon X au synchrotron SOLEIL. Des outils de dépouillement ont été mis en place afin d'analyser la microstructure, de mesurer la déformation et de caractériser les mécanismes d'endommagement à l'intérieur du matériau de façon qualitative et quantitative. Les images tomographiques fournissent des descriptions 3D sur la microstructure, qui sont les données d'entrées directes pour la simulation numérique basée sur la méthode FFT. L'utilisation des microstructures réelles permet de combiner directement les résultats de simulation avec les observations expérimentales.

## Abstract

SiC/SiC composite tubes are studied as materials for nuclear fuel cladding. A thorough understanding on the mechanisms of damage to this material requires both an experimental and a numerical study. In situ tensile tests were performed under X-ray tomography at the SOLEIL synchrotron. Post processing methods have been developed to analyze the microstructure, measure the deformation and characterize qualitatively and quantitatively the damage mechanisms inside the material. The tomographic images provide 3D descriptions on the microstructure, which are direct input data for the numerical simulation based on FFT. The use of real microstructures makes it possible to combine the simulation results directly with the experimental observations.

**Mots Clés :** Composite à matrice céramique, mécanismes d'endommagement, tomographie en rayon X, simulation FFT

## 1. Introduction

Une solution de gainage innovante est actuellement développée au CEA/DMN pour le gainage du combustible des réacteurs de quatrième génération (Génération IV) comme le GFR, Gaz Fast Reactor. La solution proposée est un tube sandwich constitué d'une couche interne de composite SiC/SiC (enroulement filamentaire), d'un liner métallique permettant d'assurer l'étanchéité du gainage, et de deux couches externes de composites SiC/SiC (tressage 2D). Le comportement mécanique de ces tubes est déterminé par les couches de composites, qui sont anisotropes, faiblement déformables (~1%) et dont le comportement macroscopique dépend fortement du choix des architectures fibreuses. Les matériaux étudiés dans cette étude sont des tubes de composites élaborés à partir d'un tressage 2D. Un des paramètres importants de ce type d'architectures fibreuses, est l'angle de tressage, qui influence à la fois la limite d'élasticité et la résistance à la rupture.

Les mécanismes de déformation de ces composites SiC/SiC sont essentiellement associés au développement de la microfissuration au sein du matériau. Des travaux de caractérisation de ces mécanismes ont été effectués par observations surfaciques [1] en utilisant la corrélation d'image numérique. En revanche, la description volumique de ces mécanismes, complémentaire des observations de surface, demeure une question ouverte. L'objectif du travail actuel est donc de mettre en place une caractérisation 3D de l'endommagement se développant dans le volume du matériau. Dans ce but, on utilise la tomographie en rayon X comme l'outil essentiel de l'approche expérimentale, et la simulation numérique par la méthode FFT comme un outil complémentaire permettant d'analyser ces résultats expérimentaux. Les outils mis en place sont utilisés pour trois angles de tressage différents afin d'examiner les effets de l'angle de tressage sur les mécanismes d'endommagement.

## 2. Essais de traction in situ sous tomographie X

### 1.1 Post-traitement des images tomographiques

Des essais de traction in situ ont été réalisés en tomographie par rayons X sur la ligne de faisceau PSICHE du synchrotron français SOLEIL. Les images tomographiques contiennent une grande richesse d'informations sur la microstructure et les mécanismes d'endommagements au sein du matériau. Toutefois, ces informations sont mêlées dans l'image 3D, dans laquelle sont également présents différents artefacts, et leur extraction nécessite la mise en place d'un post-traitement élaboré [3]. Celui-ci se compose des points suivants:

(i). Dans l'image de référence (non chargée), les micro- et macro- pores sont identifiés et distingués en fonction de leurs caractéristiques géométriques (le volume et l'allongement des micropores permet de les distinguer des macropores).

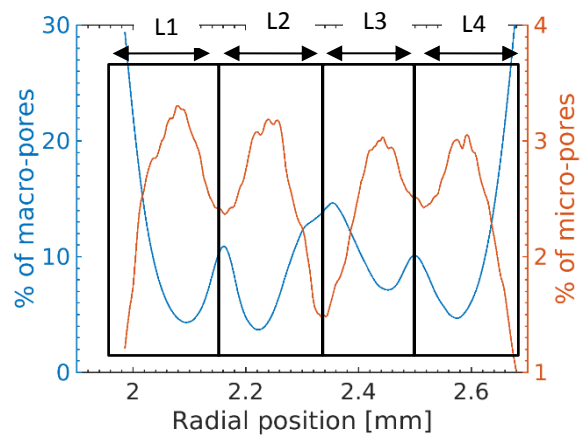

*Figure 1. Profil radial de fraction volumique de macropores et de micropores*

Les profils dans la direction radiale de macropores et de micropores sont tracés dans Fig. 1 pour le tube tressé à 45°. Une corrélation négative est observée entre les évolutions de ces deux familles de porosités. Cette observation est cohérente avec leur description qualitative : les micropores se situent au sein d'un toron, de forme allongée dans la direction du toron ; tandis que les macropores se situent entre les torons, avec des géométries assez complexes sans orientation préférentielle. Selon ces profils radiaux, l'épaisseur du tube peut être découpée en quatre sous-couches $L_i$ (voir Fig. 1), correspondant aux demi-couches de chaque couche de tressage.

(ii). Les composantes de déformation moyennes sur des couronnes concentriques sont mesurées à l'aide d'un modèle cinématique optimisé sur les champs de déplacement mesurés par corrélation d'images numériques (DVC). Leurs évolutions sur l'épaisseur du tube sont données ci-dessous (Fig. 2).

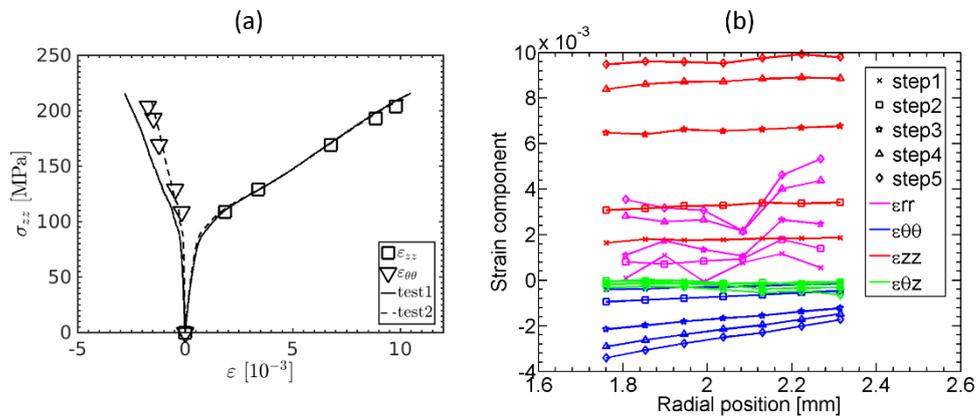

*Figure 2. (a) Composantes de déformation mesurées par DVC sur les images tomographiques, comparées aux courbes macroscopiques mesurées par extensomètres sur la surface du tube lors d'un autre essai. (b) Profils radiaux des quatre composantes de déformation mesurée par DVC.*

La légère différence entre les deux courbes de référence pour la déformation circonférentielle peut être attribuée à la réponse instable de l'extensomètre transverse positionné sur la surface rugueuse d'un tube de faible diamètre. Les déformations axiales et circonférentielles mesurées par la méthode proposée sont en bon accord avec celles mesurées par extensomètres. Ceci valide la méthode de mesure de déformations, et aussi démontre la bonne reproductibilité du comportement macroscopique des tubes étudiés.

Quatre composantes de déformation sont présentées dans la Fig. 2.b. La déformation axiale est uniforme à travers l'épaisseur. La déformation circonférentielle présente une tendance à décroitre de l'intérieur vers l'extérieur du tube. Un point surprenant est que la déformation radiale, plus oscillante, est en moyenne positive pour un chargement de traction uniaxiale.

(iii). En utilisant une technique de soustraction d'images par DVC [2], les fissures 3D induites par le chargement de traction sont extraites de la microstructure hétérogène pour chaque niveau de chargement. Ils peuvent être classés en deux familles selon leurs orientations locales: les fissures circonférentielles, perpendiculaires à la direction de traction et les fissures dans le plan, qui s'ouvrent dans l'épaisseur du tube (voir Fig. 3). L'évolution avec le chargement des fissures détectées est d'abord étudiée qualitativement par une visualisation directe en 3D des fissures détectées au sein de la microstructure.

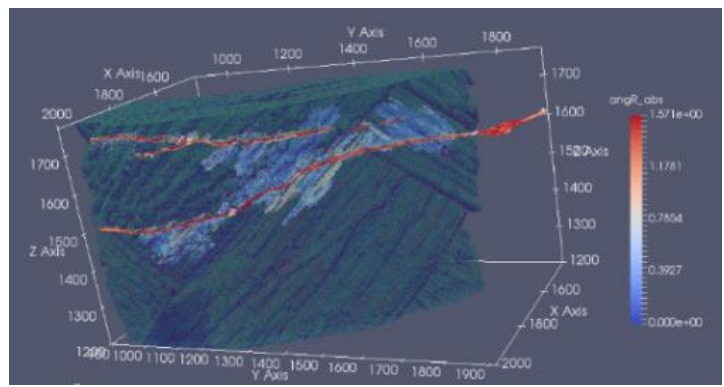

*Figure 3. Fissures classifiées superposées sur la microstructure de tressage à 45°: fissures circonférentielles (rouge) et fissures dans le plan (bleu).*

Un exemple de résultat est donné dans la Fig. 3 pour le tube tressé à 45 °. Cette figure montre les fissures détectées, colorées selon l'orientation locale et superposées sur la microstructure de tressage. La fissure dans le plan (bleue) semble être issue de la déviation des fissures circonférentielles (rouge), et sa croissance est guidée par les fibres SiC adjacentes.

(iv). Une procédure de quantification de la surface et de l'ouverture des fissures est mise en place en utilisant le niveau de gris et l'orientation locale des voxels « fissurés ». L'évolution de ces quantités est représentée en fonction du chargement appliqué à la Fig. 4. Il est notable que l'ouverture moyenne des fissures (~1.5 µm) est légèrement inférieure à la taille de voxel des images tomographiques (2.6µm), sa quantification nécessite donc une précision subvoxel.

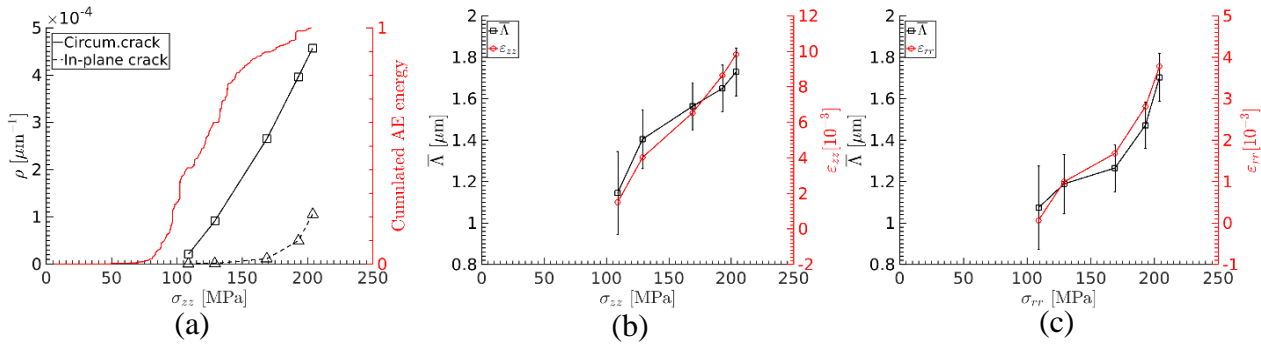

*Figure 4. (a) Evolutions de la densité surfacique des fissures circonférentielles et des fissures dans le plane, comparées à celle de l'énergie cumulée de l'émission acoustique. (b) Evolution de l'ouverture moyenne des fissures circonférentielles, comparée à celle de la déformation axiale. (c) Evolution de l'ouverture moyenne des fissures dans le plan, comparée à celle de la déformation radiale.*

La densité surfacique des fissures est définie comme le rapport entre la surface totale des fissures sur le volume de la phase solide considéré (Fig. 4.a). L'évolution de ce paramètre pour les fissures circonférentielles est quasi-linéaire jusqu'à la rupture du tube. Les fissures dans le plan commencent à apparaitre à partir de l'étape 3 (169 MPa), et leur densité surfacique augmente de plus en plus rapide jusqu'à la rupture. Contrairement à l'évolution d'énergie cumulée évaluée par Emission Acoustique, la tendance à la saturation n'est observée pour aucune de ces deux familles de fissures. Ceci implique que le développement des réseaux de fissures ne dissipe pas nécessairement beaucoup d'énergie une fois que le composite est déjà significativement endommagé.

Les ouvertures moyennes sont mesurées séparément pour les deux familles de fissures (Fig. 4.b et c). Deux changement de pentes sont présents sur l'évolution des fissures circonférentielles, alors que les fissures dans le plan s'ouvrent de plus en plus vite jusqu'à la rupture. Ces ouvertures moyennes sont des données expérimentales inédites susceptibles d'alimenter directement les modèles à base micromécanique. Leurs évolutions sont comparées à celle des déformations moyennes (macroscopiques). Cette superposition est en accord avec l'idée selon laquelle l'ouverture des fissures circonférentielles serait majoritairement responsable de la déformation axiale et laisse penser que l'ouverture des fissures dans le plan serait à l'origine de la déformation radiale positive mesurée.

**1.2 Effets de l'angle de tressage**
Les post-traitements sont implémentés dans un code MatLab, qui permet d'automatiser les dépouillements des images tomographiques du même type. Les tubes élaborés avec trois angles de tressage différents sont étudiés en utilisant les procédures présentées ci-dessus sur un tube avec un tressage à 45°.

La visualisation 3D combinée, des mécanismes d'endommagement et la microstructure complexe, envisagée pour établir les liens entre la position des fissures et la microstructure s'est avérée beaucoup trop délicate à interpréter. Ainsi, une procédure de *déroulement* a été mise en place, qui permet de transformer les informations (fissures et microstructure) du repère cylindrique au repère cartésien. Sur cette configuration *déroulée*, des projections sont effectuées au sein de chaque sous-couche (voir Fig. 1) permettant ainsi de positionner les fissures par rapport à la microstructure au sein de chaque sous-couche (Fig. 5 et 6).

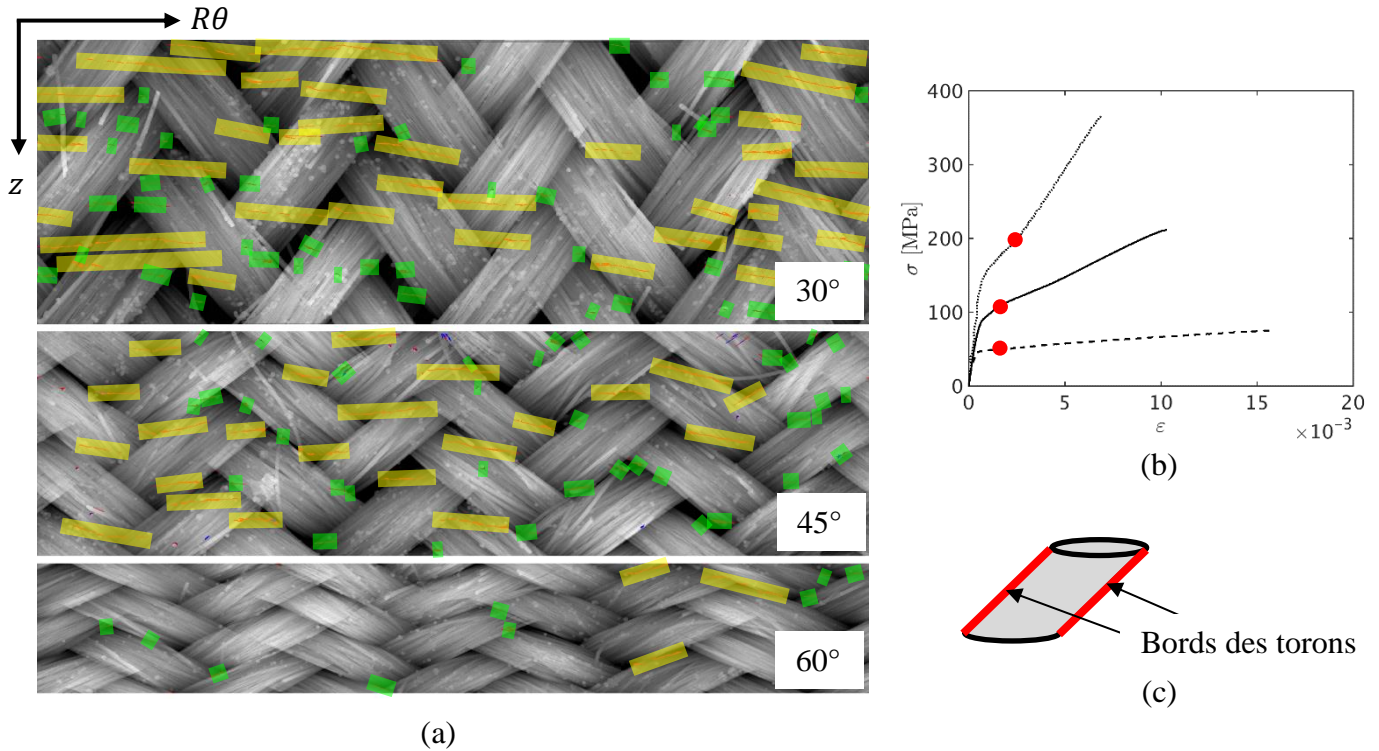

*Figure 5. (a) Fissures détectées à la première étape d'endommagement pour les trois tubes tressés à différents angles, sous-couche extérieure L4; (b) Courbes contrainte-déformation des trois tubes, avec les points rouges indiquant les niveaux de chargement correspondant aux fissures détectées dans (a) ; (c) Illustration des bords des torons*

Afin d'étudier l'initiation de fissures, on superpose les fissures détectées à la première étape de chargement à la microstructure de tressage pour chaque tube (Fig. 5). Les fissures sont mises en évidence par des barres colorées selon leurs tailles : petites fissures en vert, grandes fissures en jaune. En regardant les positions des petites fissures (vertes), il semblerait que les fissures s'amorcent préférentiellement aux bords des torons (Fig. 5.c), quel que soit angle de tressage. Cette hypothèse sera discutée à partir des champs de contraintes évalués par simulation numérique au paragraphe 3.

La propagation des fissures peut être observée par les grandes fissures dans la Fig. 6.a, qui met en évidence un effet clair de l'angle de tressage : les fissures dans le tube tressé à 60° propagent principalement entre les torons (inter-toron), alors que les fissures dans les deux tubes de faibles angles de tressage traversent les torons (intra-toron). Cette conclusion est confirmée par les fissures détectées à la dernière étape de chargement (Fig. 6.a). Dans la Fig. 6.a, les fissures dans le plan sont également tracées, leur population présente aussi un fort effet d'angle de tressage : elles sont plus nombreux dans le tube tressé à 45° que dans le 30°, et elles sont très peu propagées dans le tube tressé à 60°. Enfin, une observation plus détaillée de la fissuration dans le plan semble montrer qu'elles seraient engendrées par une déviation des fissures circonférentielles dans le « plan » du toron (Fig. 6.c).

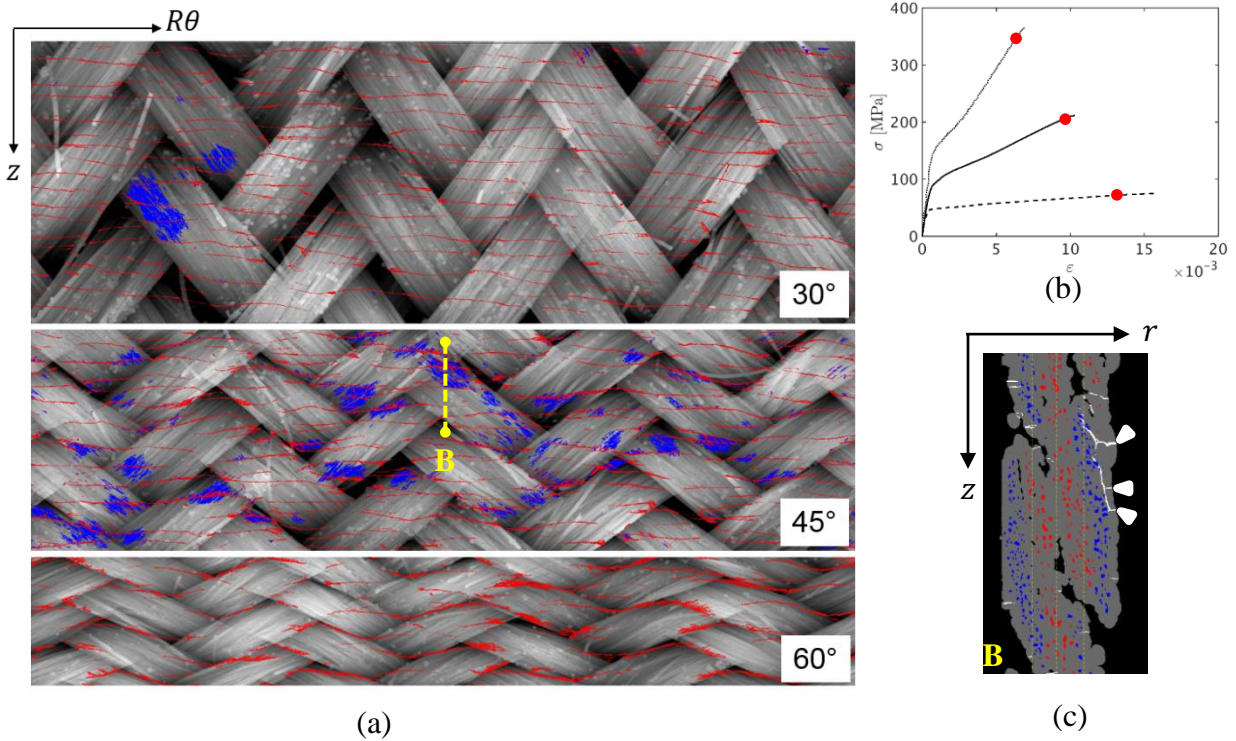

*Figure 6. (a) Fissures détectées à la dernière étape de chargement pour chaque tube testé sous-couche extérieure L4; (b) Courbes contrainte-déformation des trois tubes, avec les points rouges indiquant les niveaux de chargement correspondant aux fissures détectés dans (a) ; (c) Coupe dans le plan $rz$ selon le segment B.*

## 3. Simulation numérique par la méthode FFT

Les images tomographiques peuvent directement alimenter la simulation numérique basée sur l'algorithme de transformée de Fourrier rapide (FFT). Deux avantages principaux de cette méthode sont : (i) l'algorithme utilise une discrétisation de la microstructure sous forme d'une grille régulière permettant d'utiliser directement les images tomographiques 3D de la microstructures réelles, sans passer par une étape délicate de maillage; (ii) massivement parallélisée au sein du code AMITEX [4], la méthode peut s'appliquer à des volumes de très grande taille permettant ainsi de conserver la grande finesse de discrétisation spatiale obtenue en imagerie tomographique.

La microstructure du composite étudié est très complexe et surtout multi-échelle. Donc sa description nécessite un niveau minimum de discrétisation spatiale, afin notamment de décrire les micropores, au moins les plus importants, à l'origine de l'anisotropie élastique du toron [5]. De plus, afin de limiter les effets de bords inhérents aux conditions aux limites imposées sur les sections extrémales du tube, la longueur du tronçon simulé doit être suffisamment importante. Ces exigences imposent l'utilisation d'un code massivement parallélisé. A titre d'exemple, des tailles de cellule de 1993x1993x1685, soit environ 7 milliards de voxels, ont pu être introduites au sein du code AMITEX [3] utilisé sur le supercalculateur cobalt du TGCC (Très Grand Centre de Calcul).

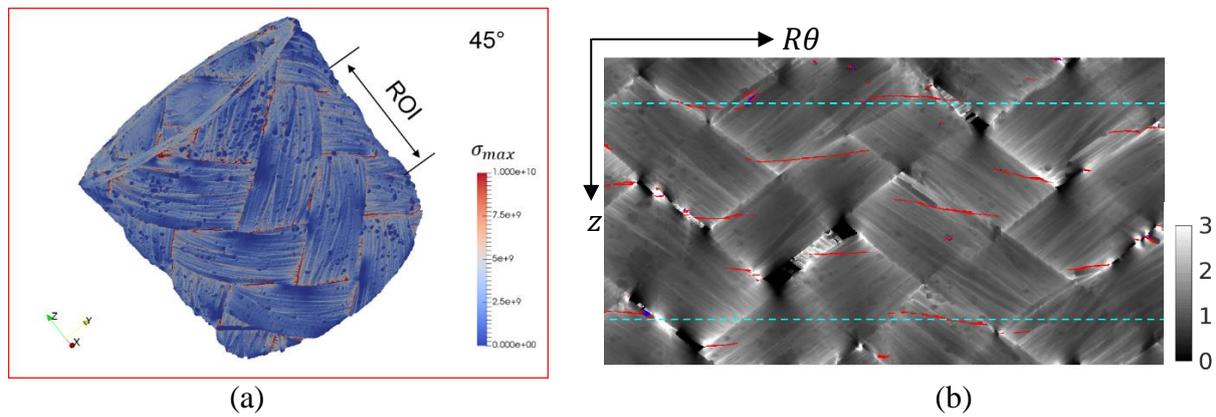

*Figure 7. (a) Distribution de contrainte principale maximale calculée par la simulation FFT pour le tube de tressage à 45° ; (b) Contrainte principale maximale moyennée dans la demi-couche extérieure selon la direction radiale, projetée dans le plan $R\theta$-z, les fissures détectés par la tomographie X sont présentées en pixels rouges*

Le résultat d'un calcul sur une microstructure non-endommagée est donnée dans la Fig. 7 pour le tube tressé à 45°. Fig. 7.a met en évidence une forte hétérogénéité, induite par la géométrie du tressage, de la distribution des contraintes principales maximales. A nouveau, afin de faciliter l'interprétation de ces champs 3D, ces contraintes sont moyennées selon la direction radiale, dans la demi-couche extérieure de tressage. Cette projection dans le plan $R\theta$-z est présentée à la Fig. 2.b, sur laquelle sont également superposées les fissures détectés par la tomographie X à la première étape d'endommagement. Ces fissures sont presque toutes reliées aux bords des torons où se situent d'importantes concentrations de contrainte. Ce dialogue direct entre la caractérisation expérimentale et la simulation numérique permet de conforter l'idée que les fissures matricielles s'initient aux bords des torons (Fig. 5.c). Cette analyse des concentrations de contrainte, en cours de finalisation, devrait également nous permettre d'expliquer l'augmentation de la limite d'élasticité avec la diminution de l'angle de tressage.

### 4. Conclusion

Pour les essais in situ sous tomographie X, un post-traitement élaboré a été mise en place, qui a permis de caractériser l'initiation et la propagation des fissures au sein des composites tressés à différents angles. Les évolutions de l'ouverture moyenne et de la surface des fissures en fonction du chargement appliqué fournissent des arguments expérimentaux pour la modélisation micromécanique. Les simulations numériques sur des microstructures réelles ont mis en évidence de fortes localisations de contraintes susceptibles d'être à l'origine de l'initiation des fissures observées expérimentalement.

**Remerciements**